\documentclass[fleqn,usenatbib,useAMS]{mnras}

\usepackage{graphicx}	
\usepackage{amsmath}	
\usepackage{amssymb}	
\usepackage{multicol}        
\usepackage{bm}		
\usepackage{pdflscape}	

\usepackage[T1]{fontenc}
\usepackage{ae,aecompl}

\usepackage{newtxtext,newtxmath}

\title[Constraining reionization with Ly$\beta$]{Constraining the second half of reionization with the Lyman-$\beta$\\forest}

\author[L.C. Keating et al.]{\parbox{\textwidth}{Laura C. Keating$^{1}$\thanks{E-mail: lkeating@cita.utoronto.ca},  Girish Kulkarni$^{2}$, Martin G. Haehnelt$^{3,4}$, Jonathan Chardin$^5$\\and Dominique Aubert$^5$}\vspace{0.4cm} \\
\parbox{\textwidth}{$^1$Canadian Institute for Theoretical Astrophysics, 60 St. George Street, University of Toronto, ON M5S 3H8, Canada\\
$^2$Department of Theoretical Physics, Tata Institute of Fundamental Research, Homi Bhabha Road, Mumbai 400005, India\\
$^3$Institute of Astronomy, University of Cambridge, Madingley Road, Cambridge CB3 0HA, UK\\
$^4$Kavli Institute for Cosmology, University of Cambridge, Madingley Road, Cambridge CB3 0HA, UK\\
$^5$Observatoire Astronomique de Strasbourg, 11 rue de l'Universite, 67000 Strasbourg, France\\}
}

\date{\today}

\pubyear{2019}

\begin{document}
\label{firstpage}
\pagerange{\pageref{firstpage}--\pageref{lastpage}}
\maketitle

\begin{abstract}
We present an analysis of the evolution of the Lyman-series forest into the epoch of reionization using cosmological radiative transfer simulations in a scenario where reionization ends late. We explore models with different midpoints of reionization and gas temperatures. We find that once the simulations have been calibrated to match the mean flux of the observed Lyman-$\alpha$ forest at $4 < z < 6$, they also naturally reproduce the distribution of effective optical depths of the Lyman-$\beta$ forest in this redshift range. We note that the tail of the largest optical depths that is most challenging to match corresponds to the long absorption trough of ULAS J0148+0600, which we have previously shown to be rare in our simulations. We consider the evolution of the Lyman-series forest out to higher redshifts, and show that future observations of the Lyman-$\beta$ forest at $z>6$ will discriminate between different reionization histories. The evolution of the Lyman-$\alpha$ and Lyman-$\gamma$ forests are less promising as a tool for pushing studies of reionization to higher redshifts due to the stronger saturation and foreground contamination, respectively.
\end{abstract}


\section{Introduction}
\label{sec:intro}

The discovery of the first quasars towards redshift 6 \citep{fan2001} provided a new probe of the high redshift Universe. By analysing the  Lyman-$\alpha$ (Ly$\alpha$) forest in the spectra of these quasars, it quickly became clear that the intergalactic medium (IGM) was evolving rapidly in this redshift range and that we were perhaps witnessing the end of the reionization epoch \citep{fan2006}. As the sample of high-quality spectra of quasars above redshift 5 has increased, so too has the quality of the constraints we can place on the properties of the IGM and on the timing of reionization \citep[see, for example,][for a review of some of the different methods that have been used]{becker2015rev}. Transmission of the Ly$\alpha$ forest, however, saturates at relatively low \ion{H}{i} fractions ($f_{\ion{H}{i}} \sim 10^{-4}$), making it increasingly difficult to use the Ly$\alpha$ forest as a probe of the IGM at redshifts pushing into the epoch of reionization.

As well as the Ly$\alpha$ forest, further information can be obtained from the spectra of these quasars by instead analysing the next line in the Lyman series, Lyman-$\beta$ (Ly$\beta$). Ly$\beta$ absorption occurs at a shorter wavelength than Ly$\alpha$ ($\lambda_{\rm{Ly}\alpha}$ = 1215.67 \AA \, \textit{vs.} $\lambda_{\rm{Ly}\beta}$ = 1025.72 \AA) and it further has a lower oscillator strength ($f_{\rm{Ly}\alpha}$ = 0.4164 \textit{vs.} $f_{\rm{Ly}\beta}$ = 0.0791). At a given point in the IGM, the optical depth to Ly$\beta$ absorption is therefore related to Ly$\alpha$ absorption by 
\begin{equation}
    \tau_{\rm{Ly}\beta} = \frac{ f_{\rm{Ly}\beta}}{f_{\rm{Ly}\alpha}} \frac{\lambda_{\rm{Ly}\beta}}{\lambda_{\rm{Ly}\alpha}} \tau_{\rm{Ly}\alpha} \approx 0.16  \, \tau_{\rm{Ly}\alpha}.
\end{equation}
In practice, the conversion factor of the \textit{effective} optical depth ($\tau_{\rm eff} = - \ln{\langle F \rangle}$, where $\langle F \rangle$ is the mean flux measured along some interval) can also depend on the density structure of the IGM \citep{oh2005,fan2006} as well as its temperature \citep{furlanetto2009}. The lower optical depth of Ly$\beta$ allows for transmission even when the \ion{H}{i} fraction of the IGM is already high enough to saturate Ly$\alpha$ transmission, and can potentially be used to push absorption line constraints on the ionization state of the IGM to higher redshifts. Indeed, \citet{barnett2017} reported a possible observation of an (unresolved) Ly$\beta$ transmission spike at $z=6.85$, a redshift much higher than the point where transmission from the Ly$\alpha$ forest has become saturated. Interpretation of the Ly$\beta$ forest is complicated, however, by the fact that the Ly$\beta$ forest is observed behind a foreground of Ly$\alpha$ forest absorption at a lower redshift 
\begin{equation}
z_{\rm{FG \, Ly}\alpha} = \frac{\lambda_{\rm{Ly}\beta}}{\lambda_{\rm{Ly}\alpha}}(1 + z_{\rm abs}) - 1,
 \end{equation}
where $z_{\rm abs}$ is the redshift of the Ly$\beta$ absorption. This separation in redshift space is large enough that the density fields can be considered to be uncorrelated \citep{dijkstra2004}, and the total observed optical depth at an individual pixel can be forward modelled in simulations by  calculating $\tau_{\beta}^{\rm obs} (z_{\rm abs}) = \tau_{\beta}(z_{\rm abs}) + \tau_{\alpha}(z_{\rm{FG \, Ly}\alpha})$, where the Ly$\beta$ optical depth at the redshift of interest ($z_{\rm abs}$) and the foreground Ly$\alpha$ optical depth at $z_{\rm{FG \, Ly}\alpha}$ are calculated for different density fields. Another possibility is to observe absorption of the next Lyman-series transition, Lyman-$\gamma$ (Ly$\gamma$), which occurs at a wavelength $\lambda_{\rm{Ly}\gamma}$ = 972.54 \AA \, and has an oscillator strength $f_{\rm{Ly}\gamma}$ = 0.029. This results in an optical depth $\tau_{\rm Ly \gamma} = 0.056 \, \tau_{\rm Ly \alpha}$.  In practice, however,  observing and interpreting the Ly$\gamma$ forest will be complicated due to the presence of foregrounds from both the Ly$\alpha$ and Ly$\beta$ forests at lower redshift.

The evolution of the effective optical depth of the Ly$\beta$ forest with redshift has long been used as a probe of the high redshift IGM \citep{lidz2002,songaila2004,fan2006}. \citet{eilers2019} recently presented a new compilation of effective optical depths of the Ly$\beta$ forest along 19 different sightlines in the redshift range $5.5 \lesssim z \lesssim 6.1$. They demonstrated that it was difficult to simultaneously reproduce the observed ratio of co-spatial Ly$\alpha$ and Ly$\beta$ optical depths in a variety of models that modelled the IGM assuming a uniform ionizing background, UV fluctuations from a varying mean free path \citep{davies2016,daloisio2018} and temperature fluctuations from inhomogeneous reionization \citep{daloisio2015,keating2018}. They found that the most successful model was one in which the temperature-density relation of the IGM was inverted, which is somewhat challenging to explain theoretically at these redshifts. If the gas was reionized early, one might expect the gas to have already cooled into the usual temperature-density relation \citep{hui1997} or, if it has been recently ionized, it may instead be close to isothermal. 

However, the models presented in \citet{eilers2019} all make the assumption that the IGM is completely ionized above redshift 5.5. \citet{kulkarni2019} used cosmological radiative transfer simulations to show that the broad distribution of Ly$\alpha$ forest opacities along different lines of sight \citep{becker2015,bosman2018,eilers2018} can be explained by a model where reionization ends late, with islands of neutral hydrogen persisting down to $z \sim 5.3$. Such a scenario was also proposed by  \citet{lidz2007} and \citet{mesinger2010}. \citet{keating2019} further showed that a late-end reionization model also explained the observed long absorption trough in the spectrum of ULAS J0148+0666 \citep{becker2015}, as well as its anti-correlation with Ly$\alpha$ emitters \citep{becker2018}. This result was subsequently demonstrated using a semi-numeric method in \citet{nasir2019}.

In this paper, we present a study of the Ly$\alpha$, Ly$\beta$ and Ly$\gamma$ forests in late-end reionization models, using cosmological radiative transfer simulations carefully calibrated to match the properties of the observed Ly$\alpha$ forest. In Section \ref{sec:sims}, we describe the simulations we analyse here. We compare these simulations to the existing observations of \citet{eilers2019} in Section \ref{sec:cdfs}, and make predictions for future observations in Section  \ref{sec:highz}. Finally, in Section \ref{sec:conclusions}, we make our conclusions.

\begin{figure}
\includegraphics[width=\columnwidth]{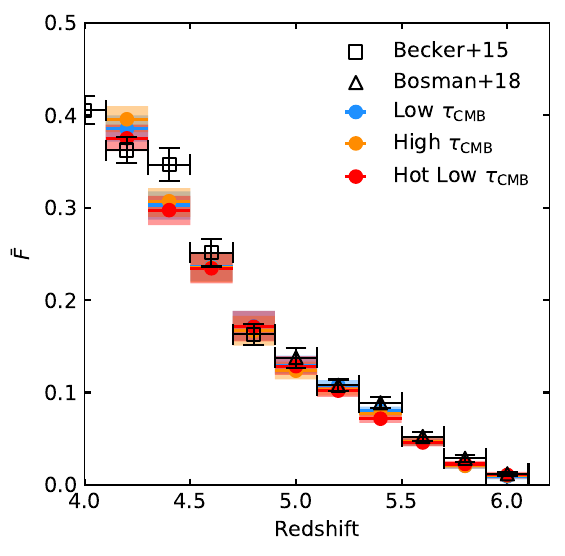}
    \caption{The evolution of the mean flux with redshift measured in our radiative transfer simulations and compared with observations from \citet{becker2015} and \citet{bosman2018}.}
    \label{fig:meanflux}
\end{figure}

\section{Cosmological radiative transfer simulations of the intergalactic medium}
\label{sec:sims}

We model the ionization state of the IGM by post-processing cosmological hydrodynamic simulations with the radiative transfer code \textsc{aton} \citep{aubert2008,aubert2010}. The underlying hydrodynamic simulation was performed with the Tree-PM SPH code \textsc{p-gadget-3} \citep[last described in][]{springel2005}. The simulation volume has box size 160 Mpc/$h$ and was run with 2048$^{3}$ gas particles, resulting in gas particle masses $m_{\rm gas} = 6.4 \times 10^{6} \, M_{\odot}/h$. We used a gravitational softening length $l_{\rm soft} = 3.1$ kpc/$h$. The initial conditions were taken from a simulation performed as part of the Sherwood simulation suite \citep{bolton2017} and were generated at $z=99$. The cosmological parameters used were $\Omega_{\rm m} = 0.308, \, \Omega_{\Lambda} = 0.692, \, h=0.678, \, \Omega_{\rm b} = 0.0482, \, \sigma_{8} = 0.829$ and $n_{\rm s} =0.961$ \citep{planck2014}. The simulation uses a simplified and computationally efficient prescription for star formation, which turns any gas with density 1000 times the mean density of the Universe and temperature $T < 10^{5}$ K into collisionless star particles. Removing this cold and dense gas from the simulation reduces the time required to perform the simulation, but does not have a significant effect on the low density gas making up the IGM \citep{viel2004}. As we perform the radiative transfer in post-processing, we still wish to account for the pressure smoothing of the density field and so include the effect of photoheating in the hydrodynamical simulation by imposing a uniform UV background \citep{haardtmadau2012}.

\begin{figure*}
\includegraphics[width=2\columnwidth]{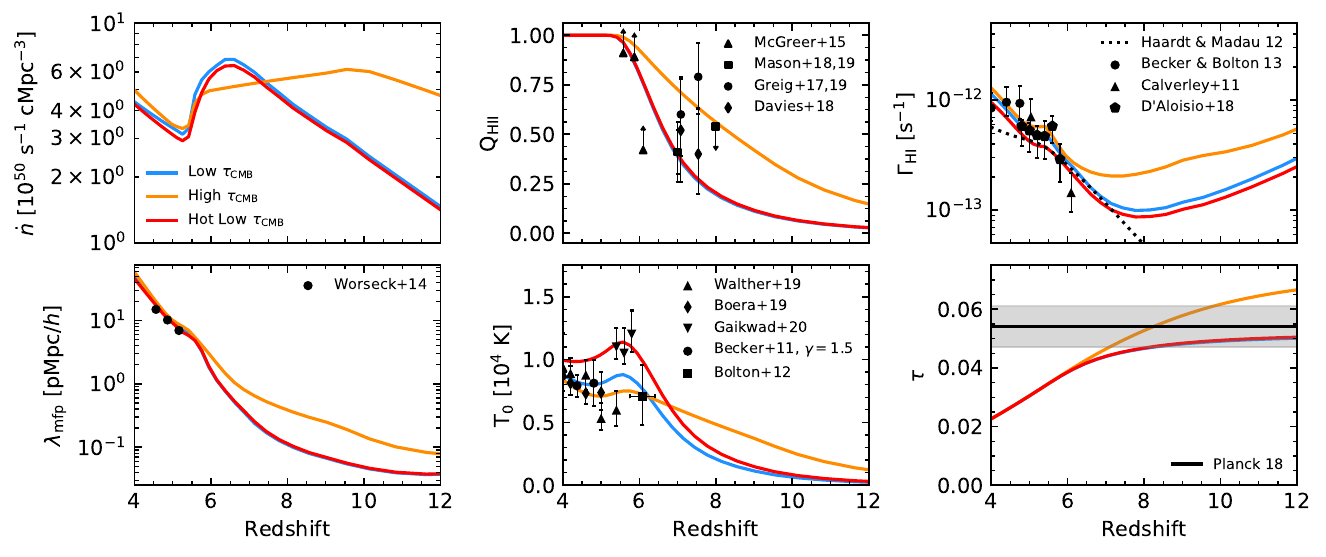}
    \caption{Top left: Evolution of the volume ionizing emissivity ($\dot{n}$) with redshift in the three simulations we present. Top middle: Evolution of the ionized hydrogen fraction ($Q_{\ion{H}{ii}}$) with redshift compared to constraints from the dark pixel fraction \citep{mcgreer2015}, the fraction of continuum-selected galaxies showing Ly$\alpha$ emission \citep{mason2018,mason2019} and analyses of the strength of the damping wings observed in $z >7$ quasars \citep{greig2017,greig2019,davies2018}. Top right: Evolution of the photoionization rate ($\Gamma_{\ion{H}{i}}$) with redshift in ionized regions in the simulations. Also shown is the \citet{haardtmadau2012} model for the UV background used in the underlying hydrodynamic simulation. We compare with observational constraints from \citet{becker2013}, \citet{calverley2011} and \citet{daloisio2018}. Bottom left: Evolution of the mean free path ($\lambda_{\rm mfp}$) at 912 \AA \, with redshift in the simulations compared to the observations of \citet{worseck2014}. Bottom middle: Evolution of the volume-weighted mean temperature at the mean density of the Universe ($T_{0}$) with redshift in the simulations. Also shown are measurements from \citet{walther2019}, \citet{boera2019}, \citet{gaikwad2020}, \citet{becker2011} and \citet{bolton2012}. Bottom right: Evolution of the optical depth to Thomson scattering of CMB photons ($\tau_{\rm CMB}$) with redshift. Also shown is the constraint from \citet{planck2018}.}
    \label{fig:history}
\end{figure*}

\begin{table}
	\centering
	\caption{Details of the simulations presented in this paper. For each simulation we give the photon energy ($E_{\gamma}$), the redshift at which 50 per cent of the gas is ionized ($z_{50}$), the redshift at which 99.9 per cent of the gas is ionized ($z_{99.9}$) and the optical depth to Thomson scattering of CMB photons ($\tau_{\rm CMB}$).}
	\begin{tabular}{ccccc} 
		\hline
		Label & $E_{\gamma}$ (eV) & $z_{50}$ & $z_{99.9}$ & $\tau_{\rm CMB}$ \\
		\hline
		Low-$\tau_{\rm CMB}$ & 17.1 & 6.7 & 5.2 & 0.051\\
		High-$\tau_{\rm CMB}$ & 17.1 & 8.4 & 5.3 & 0.071\\
		Hot Low-$\tau_{\rm CMB}$ & 18.6 & 6.7 & 5.2 & 0.051\\
		\hline
	\end{tabular}
	 \label{table:sims}
\end{table}

We use outputs from this hydrodynamical simulation every 40 Myr starting at $z=19.5$ and down to $z=4$ as the input density fields for our radiative transfer calculation. As \textsc{aton} requires cartesian grids as inputs, we map the SPH particles onto $2048^{3}$ grids (with cellsize $\Delta x = 78.125$ kpc/$h$) using a cubic spline interpolation scheme. To assign sources we follow the method outlined in \citet{chardin2015}. Using the halo catalogues generated from the simulation, we assign sources to all haloes with $M_{\rm halo} > 10^{9} \, M_{\odot}/h$. At each redshift, we assume a total volume emissivity (described in more detail below). This total emissivity is divided up among all the sources and weighted by the mass of the host halo, such that the ionizing emissivity of each source is directly proportional to the mass of the host halo \citep[see also][]{iliev2006}.

\textsc{aton} solves the radiative transfer equation using a moment-based method with the M1 closure relation for the Eddington tensor \citep{levermore1984J,gnedin2001,aubert2008}. \textsc{aton}  makes use of GPU acceleration, which significantly reduces the time required to perform a simulation and allows us to use the full speed of light in the simulations. We perform the radiative transfer using a single frequency band, with appropriate photon energies and photoionization cross sections calculated using the optically thick grey approximation \citep[see, e.g.,][]{pawlik2011} and assume that the input spectrum is a blackbody. We present three different radiative transfer simulations here, described in Table \ref{table:sims}. We present models with two different ionization histories (the Low-$\tau_{\rm CMB}$ and High-$\tau_{\rm CMB}$ models) that use the same ionizing photon energy, and models with different ionizing photon energies but similar ionization histories (the Low-$\tau_{\rm CMB}$ and Hot Low-$\tau_{\rm CMB}$ models). We use two different input blackbody temperatures: $T=30000$ K \citep[as in][]{keating2019} and $T=40000$ K (as in \citealt{gaikwad2020} and Puchwein et al., in prep.). For the $T=30000$ K blackbody, this results in a photon energy $E_{\gamma} = 17.1$ eV and a photoionization cross section $\sigma_{\gamma} = 3.9 \times 10^{-18}$ cm$^{2}$. For the $T=40000$ K blackbody, this results in a photon energy $E_{\gamma} = 18.6$ eV and a photoionization cross section $\sigma_{\gamma} = 3.4 \times 10^{-18}$ cm$^{2}$. As in \citet{keating2019}, we further account for a rising temperature at $z < 5.2$ by increasing the photon energy linearly with the cosmic scale factor in the simulations, such that it reaches 23.8 eV by $z=4$. This is intended to mimic the effect of the photoheating of the IGM during \ion{He}{ii} reionization and helps us to match the mean flux of the Ly$\alpha$ forest approaching redshift 4.

We choose the ionizing emissivity in the simulations by aiming to match the observed mean flux of the Ly$\alpha$ forest \citep{becker2015,bosman2018}. We measure this mean flux from spectra constructed along sightlines taken from a lightcone extracted from the simulation on the fly, to allow us to capture the rapid evolution in the IGM towards the end of reionization. A comparison between the mean flux measured in our simulations and the observations is shown in Figure \ref{fig:meanflux}. Due to the short time required to perform a simulation with \textsc{aton}, we can run many models and modify the emissivity as required until we reach good agreement with the observations. The starting point for our Low-$\tau_{\rm CMB}$ and Hot Low-$\tau_{\rm CMB}$ models is the input emissivity of the \citet{puchwein2019} synthesis model of the UV background. As demonstrated in \citet{kulkarni2019}, this produces an ionization history that is in good agreement with the measurements of the CMB optical depth to Thomson scattering \citep{planck2018}. We further present here a model which uses the emissivity model of the \citet{haardtmadau2012} UV background as a starting point for our High-$\tau_{\rm CMB}$ model, which results in more ionizing photons at high redshift and shifts the midpoint of reionization to a higher redshift (and hence increases the $\tau_{\rm CMB}$). 

We present the resulting emissivity evolution for the three models in the top left panel of Figure \ref{fig:history}. All models require a rather rapidly declining emissivity towards the end of reionization. Note that this should be considered as an ``effective'' ionizing emissivity required to match the observed mean flux, as our simulations do not properly resolve the interstellar medium and escape of ionizing photons from the host galaxies.  As shown here, using a different ionizing spectrum requires a small change in the shape of the emissivity evolution (due to the temperature dependence of the recombination rate, which changes the \ion{H}{i} fraction in the ionized IGM and hence the mean flux). As in \cite{keating2019}, we find that matching the mean flux of the Ly$\alpha$ forest at $z \lesssim 5.5$ requires a emissivity that increases with decreasing redshift. This could perhaps be explained by an increasing contribution of AGN to the UVB towards lower redshift, raising the total emissivity of ionizing photons. This motivates our choice to increase the photon energy in the simulation as the harder photons emitted from AGN will ionize \ion{He}{ii} and heat the gas. A more detailed account of the evolution relative contributions of galaxies and AGN with redshift would require more careful modelling of multiple populations of sources, and likely a larger volume simulation than presented here. For these reasons, and the reasons discussed above, we do not make any strong conclusions on the redshift evolution of our emissivity, and just note that using this redshift evolution results in synthetic spectra that match the observed mean flux quite well over the redshift range considered here.

As well as matching the mean flux constraints, we also compare these models with a variety of measured properties of the IGM above redshift 4 (Figure \ref{fig:history}). The redshift evolution of the \ion{H}{i} fraction in our models roughly spans the allowed range from the Ly$\alpha$ absorption constraints (both from Ly$\alpha$ forest statistics and Ly$\alpha$ emission from galaxies). As expected, the Low-$\tau_{\rm CMB}$ and Hot Low-$\tau_{\rm CMB}$ models come in at the lower end of this range, with a midpoint of reionization of $z=6.7$ in both models. These models also have optical depths to Thomson scattering in good agreement with the results from the $Planck$ satellite. In contrast, the High-$\tau_{\rm CMB}$ model has a higher midpoint of reionization ($z=8.4$) and it is disfavoured by the CMB results \citep{planck2018}, although it is just about allowed by the Ly$\alpha$ absorption/attenuation constraints and we therefore include it to explore the sensitivity of the Ly$\alpha$ and Ly$\beta$ forests to different reionization histories.

The temperature of the gas in these models depends both on the reionization history and the energy of ionizing photons we assume in the simulation. We show the evolution of the volume-weighted mean temperature at mean density in the bottom middle panel of Figure \ref{fig:history}. The models with a lower photon energy (Low-$\tau_{\rm CMB}$ and High-$\tau_{\rm CMB}$) are in reasonable agreement with a range of temperature measurements of the high-redshift IGM, although they do fall below the recent measurements of the IGM temperature from the widths of transmission spikes in high resolution quasar spectra \citep{gaikwad2020}. These transmission spikes probe low density regions, and seem to prefer higher temperatures than those inferred from analyses of the flux power spectrum \citep[e.g.,][]{boera2019,walther2019}. We remain agnostic as to the actual temperature of the IGM and simply also present a hotter model (Hot Low-$\tau_{\rm CMB}$). We also note that the resolution of these simulations is insufficient to capture the widths of ionization fronts required to accurately model the IGM temperature \citep{daloisio2019}, so the photon energies we require to match the observed temperature constraints are probably overestimates. We find that all three models are in good agreement with estimates of the \ion{H}{i} photoionization rate and mean free path at 912 \AA.

\begin{figure}
\includegraphics[width=\columnwidth]{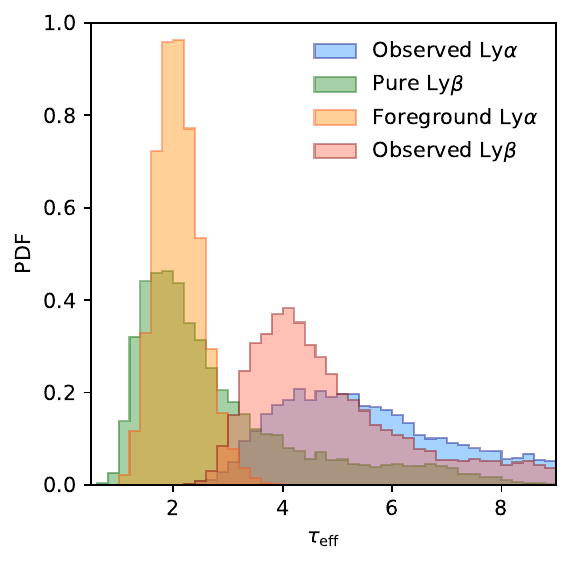}
    \caption{Distribution of effective optical depths in the Low-$\tau_{\rm CMB}$ model in the redshift range $5.9 < z < 6.1$. The foreground Ly$\alpha$ comes from the redshift range $4.8 \lesssim z \lesssim 5$. We show the observed Ly$\alpha$ effective optical depths (blue), the pure Ly$\beta$ effective optical depths (green), the foreground Ly$\alpha$ effective optical depths (orange) and the observed Ly$\beta$ effective optical depths (red).}
    \label{fig:taudist}
\end{figure}

\section{Joint analysis of the Lyman-$\alpha$ and Lyman-$\beta$ forests in a late-end reionization model} 
\label{sec:cdfs}

We next examine the evolution of the Ly$\alpha$ and Ly$\beta$ forests in these simulations. We construct synthetic spectra along lines of sight taken through the lightcone we extracted from the simulations on the fly. The optical depth is calculated using the analytic approximation to a Voigt profile presented in \citet{teppergarcia2006}. Following \citet{eilers2019}, we measure the effective optical depth along skewers with length 40 Mpc. As our simulations contain significant amounts of neutral hydrogen in the redshift range where we are comparing the models against observations, we calculate the Ly$\beta$ optical depth directly to properly account for the effect of the weaker damping wings \citep[e.g.,][]{malloy2015}, rather than simply rescaling by the oscillator strengths and wavelengths as described in Section \ref{sec:intro} and \citet{eilers2019}.  As discussed in Section \ref{sec:intro}, when comparing the Ly$\beta$ forest in our simulations against the observations, we must also account for the contribution of the foreground Ly$\alpha$ absorption. Following \citet{eilers2019}, we show how this changes the observed Ly$\beta$ effective optical depth in Figure \ref{fig:taudist} for the Low-$\tau_{\rm CMB}$ model (but the results are similar for our High-$\tau_{\rm CMB}$ and Hot Low-$\tau_{\rm CMB}$ models). As expected, accounting for the foreground Ly$\alpha$ absorption shifts the observed Ly$\beta$ effective optical depths to higher values. We note that one difference between our simulations and the equivalent figure shown in \citet{eilers2019} is the long tail towards high effective optical depths for the pure Ly$\beta$ case (and hence also the observed  Ly$\beta$ case) that is present in our simulations. This is due to the late-end reionization models we present here, with the darkest sightlines in our models corresponding to regions that contain residual islands of neutral hydrogen below redshift 6.

\begin{figure*}
\includegraphics[width=2\columnwidth]{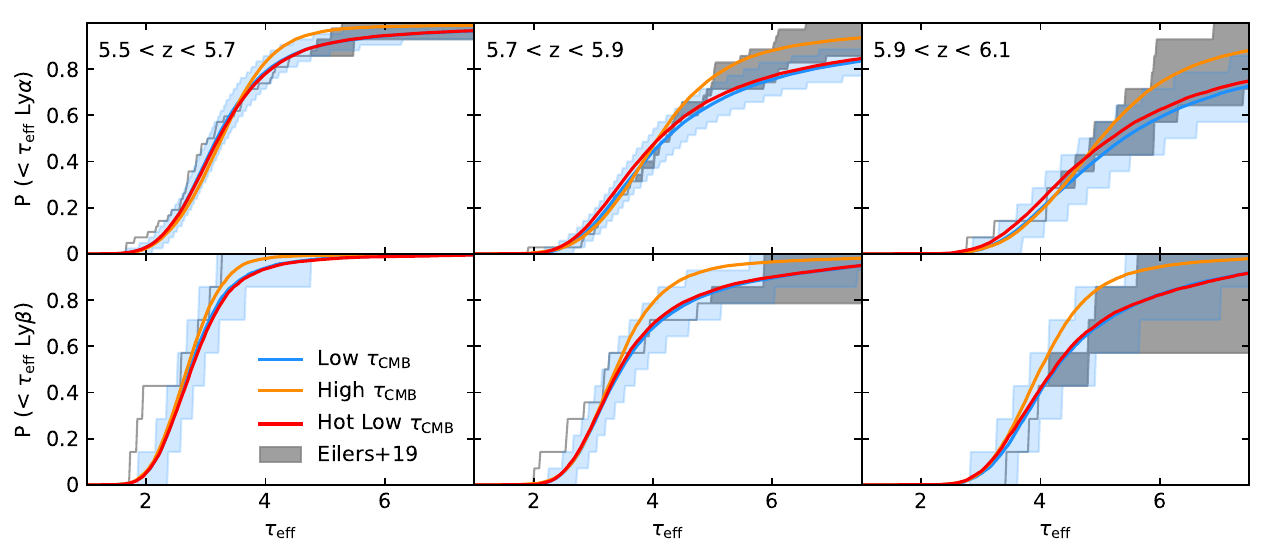}
    \caption{Cumulative distribution of the effective optical depth of the Ly$\alpha$ forest (top row) and Ly$\beta$ forest (bottom row) measured along skewers of length 40 Mpc. We show three different redshift intervals: $5.5 < z < 5.7$ (left), $5.7 < z < 5.9$ (middle) and $5.9 < z < 6.1$ (right). The coloured lines show the median CDFs for each model (blue for Low-$\tau_{\rm CMB}$ model, orange for High-$\tau_{\rm CMB}$ and red for Hot Low-$\tau_{\rm CMB}$). The blue shaded region accounts for the effect of cosmic variance in the Low-$\tau_{\rm CMB}$ model: the shaded region spans the 15$^{\rm th}$ and 85$^{\rm th}$ percentile range for 1000 CDFs calculated using the same number of sightlines as the \citet{eilers2019} compilation. The strength of this effect is similar for all models. The grey shaded regions show the cumulative distributions calculated from the \citet{eilers2019} sample. Upper limits are taken into account by showing both the case where the mean flux is equal to twice the noise (upper boundary), or assuming that the detected mean flux is correct (lower boundary). In cases where the observed mean flux is negative the sightlines are assumed to have $\tau_{\rm eff} > 8$. }
    \label{fig:cdfs}
\end{figure*}

We compare the cumulative distribution functions (CDFs) of Ly$\alpha$ and Ly$\beta$ effective optical depths calculated in our simulations to the \citet{eilers2019} measurements, shown in the top row of Figure \ref{fig:cdfs}. Looking at the Ly$\alpha$ distributions, the first thing to note is that although the simulations were calibrated to match the \citet{bosman2018} measurements of the mean flux in this redshift range, the agreement with the \citet{eilers2019} data is also quite good. This may suggest that different groups are now converging on a consensus value for the mean flux at high redshift. Second, as discussed in \citet{eilers2019}, accounting for noise in the mock observations can decrease the width of the distributions. We have not added noise to our spectra here, but instead show ``optimistic'' and ``pessimistic'' cases for the observations as in \citet{nasir2019}. The optimistic case shows the case where upper limits are treated as detections. The pessimistic case instead takes the observed value for the mean flux (even if is below twice the estimate of the noise level). In cases where the observed flux is negative we assume the sightlines are completely dark. We note that for most models, our distributions fall below the CDFs calculated from the upper limits, with long tails towards high effective optical depths. Therefore adding the noise would bring the models into closer agreement with the CDFs constructed treating upper limits as detections. Comparing the differences between the Ly$\alpha$ CDFs of the models, we find that although the three simulations we present have nearly identical mean flux, their effective optical depth distributions are somewhat different. The Low-$\tau_{\rm CMB}$ and Hot Low-$\tau_{\rm CMB}$ models have CDFs that are slightly broader than the High-$\tau_{\rm CMB}$ model, with tails of more opaque sightlines. This difference becomes more apparent with increasing redshift, and the implications of this will be discussed in Section \ref{sec:highz}. We also note that as the observations probe a limited number of sightlines, the effect of cosmic variance can play a role, and we estimate the magnitude of this effect for the Low-$\tau_{\rm CMB}$ model in the blue shaded region of Figure \ref{fig:cdfs}.

We next compare our models to the CDFs of Ly$\beta$ effective optical depth against the observations of \citet{eilers2019}, shown in the bottom row of Figure \ref{fig:cdfs}. We find that, in contrast to the models presented in \citet{eilers2019}, our models do a reasonable job of matching the observed CDFs. There are however some discrepancies, most notably the most transmissive sightlines at $5.5 < z< 5.7$ which are not reproduced within the 15$^{\rm th}$/85$^{\rm th}$ percentile range of our mock CDFs. However, we do find sightlines such as this if we extend this range out to the 5$^{\rm th}$/95$^{\rm th}$ percentile range. Unlike \citet{nasir2019}, who also investigated the evolution of the Ly$\beta$ forest in a late-end reionization model, we do find that our models produce dark Ly$\beta$ sightlines in the redshift range $5.7 < z< 5.9$. This is likely due to differences in modelling the ionization state of the gas introduced between the semi-numeric method they employ and the radiative transfer simulations presented here. Furthermore, unlike \citet{nasir2019}, we do not perform any rescaling of the photoionization rate to match the mean flux of the Ly$\alpha$ forest. This means that the reionization history (and mean free path) in our models is completely consistent with the mean flux of our synthetic spectra. The simulations presented here also resolve the ionization state of the gas at a factor 10 higher spatial resolution than in \citet{nasir2019}. We do not find any substantial differences between our three models, but again note that the High-$\tau_{\rm CMB}$ model produces comparatively fewer dark sightlines than the Low-$\tau_{\rm CMB}$ and Hot Low-$\tau_{\rm CMB}$ models towards high redshift.

\begin{figure*}
\includegraphics[width=2\columnwidth]{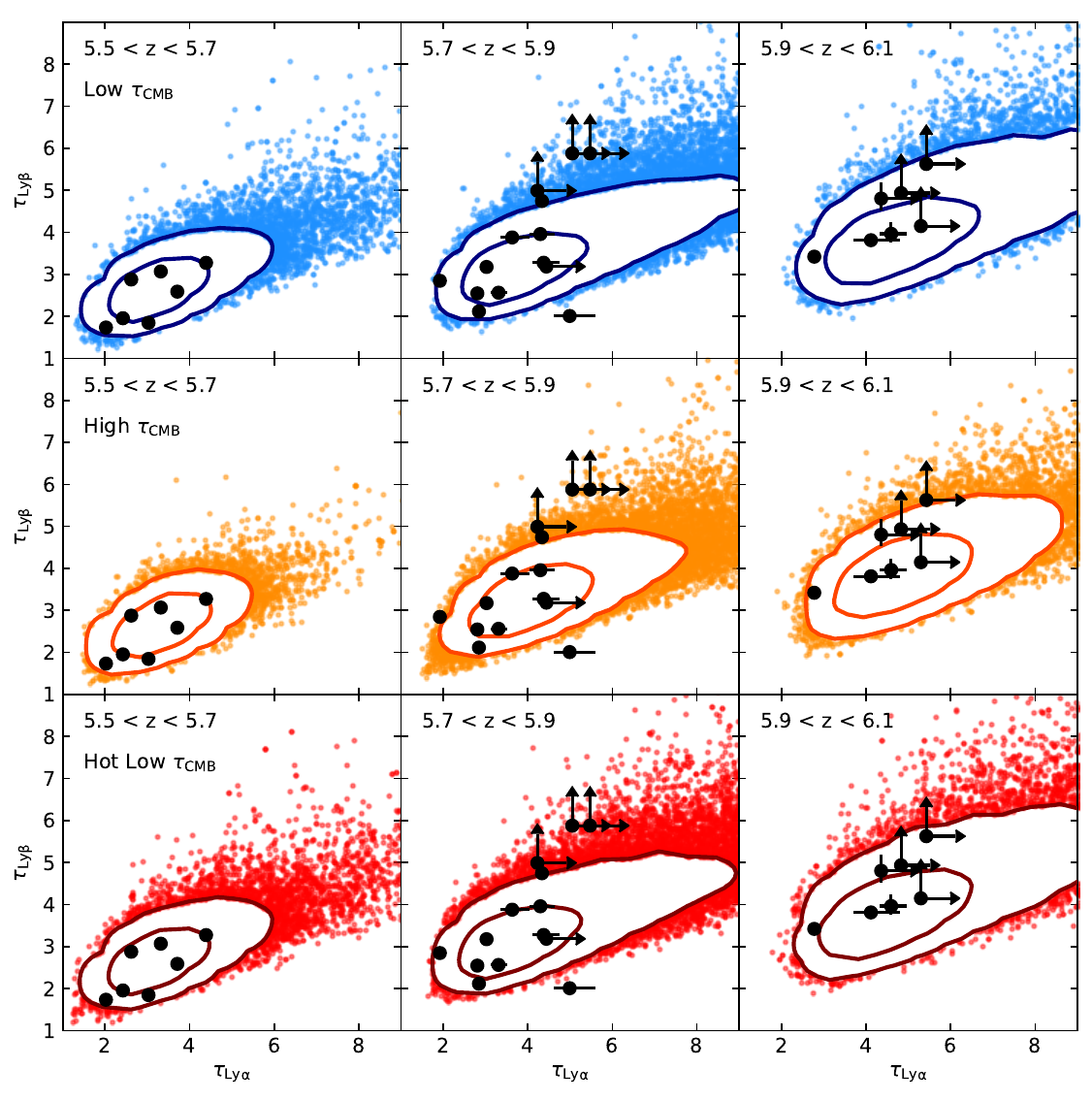}
    \caption{Relation between the Ly$\alpha$ and Ly$\beta$ effective optical depths measured in our three models: Low-$\tau_{\rm CMB}$ (top), High-$\tau_{\rm CMB}$ (middle) and Hot Low-$\tau_{\rm CMB}$ (bottom). The contours show the region enclosing 68 per cent and 95 per cent of the points. Outside these contours the coloured points represent individual sightlines. The three columns show three different redshift intervals: $6.1 < z < 6.3$ (left), $6.3 < z < 6.5$ (middle) and $6.5 < z < 6.7$ (right). The black points are taken from \citet{eilers2019}.}
    \label{fig:contours}
\end{figure*}

We next investigate the redshift evolution of co-spatial measurements of the Ly$\alpha$ and Ly$\beta$ effective optical depths. The results for the three models in the three redshift bins are shown in Figure \ref{fig:contours}. As in Figure \ref{fig:cdfs}, we do not find large differences between our three models, although as discussed above, the Low-$\tau_{\rm CMB}$ and Hot Low-$\tau_{\rm CMB}$ models predict a higher incidence rate of opaque sightlines. We compare with the data from \citet{eilers2019}. Again, we have not added noise to the mock spectra as this is already accounted for in the error bars and upper limits measured by \citet{eilers2019}. We also do not account for the uncertainties in fitting the continuua of these spectra, as this was shown in \citet{eilers2019} to only have a small effect on the scatter of the distribution of the Ly$\alpha$ and Ly$\beta$ effective optical depths. We plot contours enclosing 68 per cent and 98 per cent of our sightlines by calculating a 2D kernel density estimate of the data using a gaussian kernel with bandwidth equal to 0.2. Outside these contours we plot the results from individual sightlines. We find that the results are generally in quite good agreement, with all the observations in the $5.5<z<5.7$ and $5.9<z<6.1$ bins lying within or very close to the 95 per cent confidence interval of our models.

\begin{figure*}
\includegraphics[width=2\columnwidth]{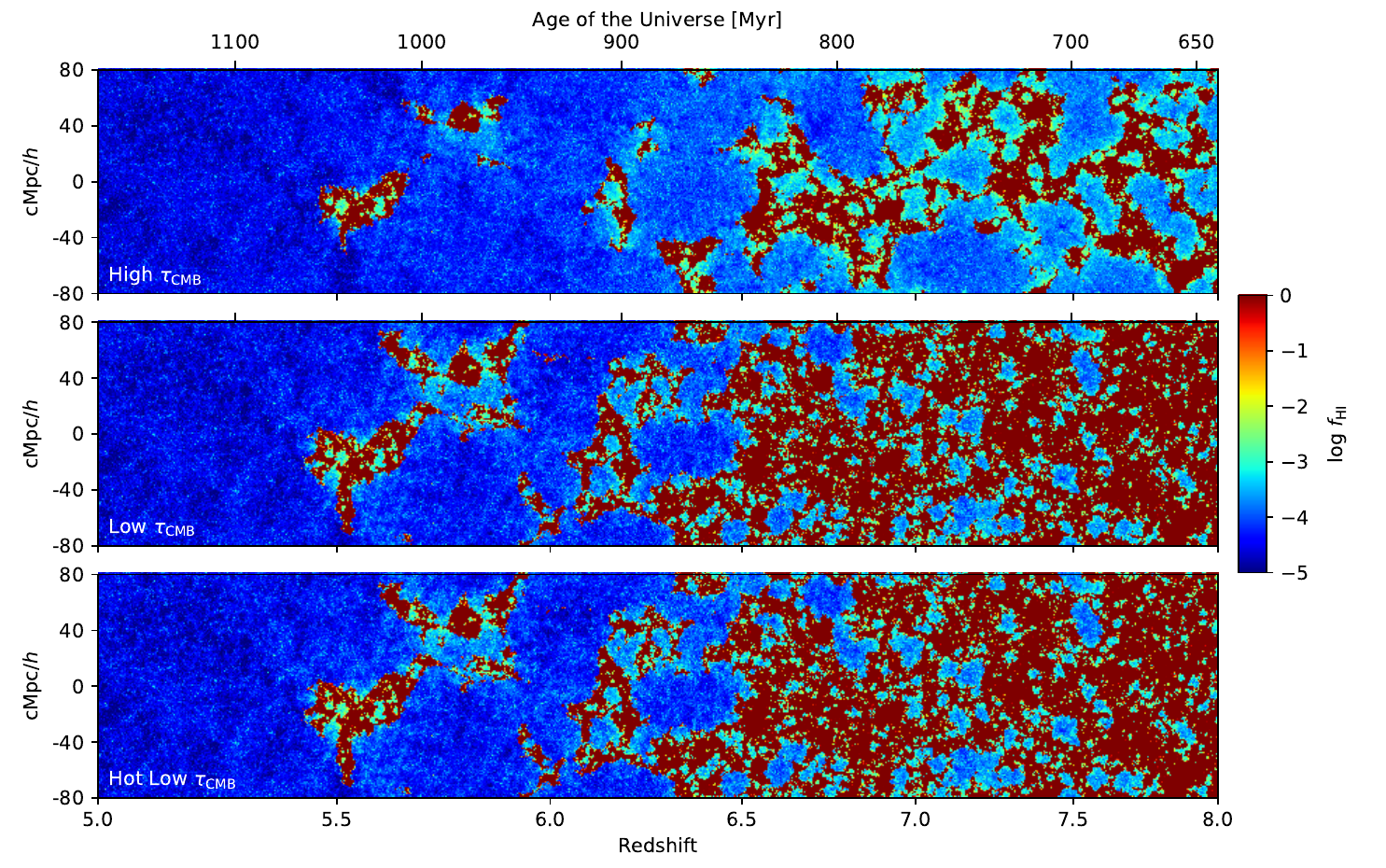}
    \caption{Lightcones showing the evolution of the \ion{H}{i} fraction with redshift for the three models presented here. The top panel shows the High-$\tau_{\rm CMB}$ model, the middle panel shows the Low-$\tau_{\rm CMB}$ model and the bottom panel shows the Hot Low-$\tau_{\rm CMB}$ model.}
    \label{fig:lc}
\end{figure*}

As also pointed out in \citet{nasir2019}, we find that the largest inconsistency between the models and the observations occurs in the $5.7<z<5.9$ bin. We note however that the most extreme points, with $\tau_{\rm eff, Ly\beta} \sim 6$ in that redshift bin, are measurements from the sightline ULAS J0148+0600 which is the sightline that hosts the extreme 110 Mpc/$h$ absorption trough identified in \citet{becker2015}. We previously showed in \citet{keating2019} that sightlines as extreme as this were uncommon in our simulation (perhaps due to the limited volume of the simulation box) and this has also been demonstrated in \citet{nasir2019}. It is therefore unsurprising that these points lie outside of the 95 per cent confidence interval calculated for our models. We do note though that there are a handful of sightlines in all three models we present that have Ly$\alpha$ and Ly$\beta$ effective optical depths high enough to be consistent with the observations. We note that reproducing converged results in the ratio of Ly$\alpha$ to Ly$\beta$ effective optical depths may require hydrodynamical simulations with higher mass resolution than the one we utilise here (see, e.g., the appendices of \citealt{becker2015} and \citealt{eilers2019}). The simulation presented here represents our best compromise between mass resolution and a volume large enough to capture the process of reionization.

The late-end reionization models we present here are consistent with the observed effective optical depths of the Ly$\alpha$ and Ly$\beta$ forests above $z=5.5$. The current data do not however differentiate between models with different midpoints of reionization or gas temperatures, as all of our models show similar results. There does not appear to be a need to invoke an inverted temperature-density relation to explain the observations, as in \citet{eilers2019}. However, the most extreme sightlines have a low incidence rate in the simulations presented here and depend on the details of the modelling \citep[see][]{nasir2019}. Further theoretical work in larger simulation volumes, in tandem with a search for more extreme absorption troughs in high quality spectra above $z=5.5$, is therefore required to understand whether the incidence rate of these sightlines may require an even later end to reionization.

\section{Predictions for future observations}
\label{sec:highz}

\begin{figure*}
\includegraphics[width=2\columnwidth]{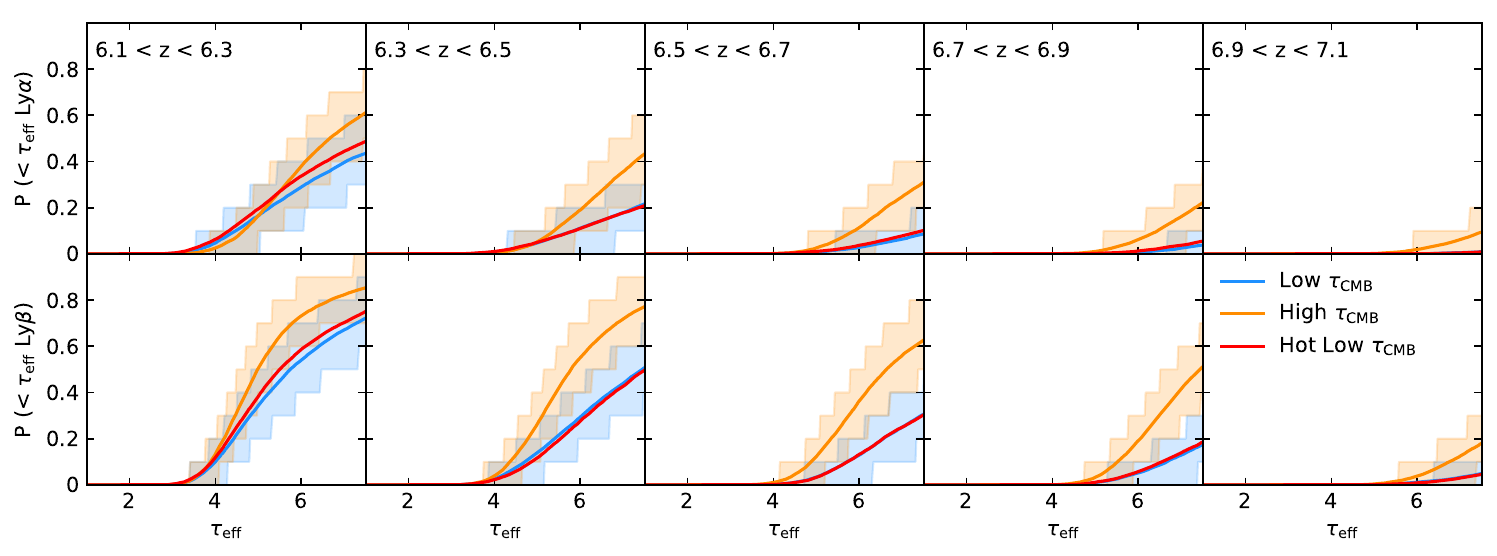}
    \caption{Cumulative distribution of the effective optical depth of the Ly$\alpha$ forest (top row) and Ly$\beta$ forest (bottom row) measured along skewers of length 40 Mpc. We show five different redshift intervals. From left to right: $6.1 < z < 6.3$, $6.3 < z < 6.5$, $6.5 < z < 6.7$, $6.7 < z < 6.9$ and $6.9 < z < 7.1$. The coloured lines show the median CDFs for each model (blue for Low-$\tau_{\rm CMB}$ model, orange for High-$\tau_{\rm CMB}$ and red for Hot Low-$\tau_{\rm CMB}$). The blue and yellow shaded regions account for the effect of cosmic variance in the in the Low-$\tau_{\rm CMB}$ model and High-$\tau_{\rm CMB}$ models respectively: the shaded region spans the 15$^{\rm th}$ and 85$^{\rm th}$ percentile range for 1000 CDFs calculated assuming a future compilation of 10 measurements in each bin. The strength of this effect is similar for the Hot Low-$\tau_{\rm CMB}$ model.}
    \label{fig:highz}
\end{figure*}

The number of high-redshift quasars identified in large surveys continues to grow, with three published quasars with redshifts above 7 to date \citep{mortlock2011,banados2018,wang2018} and many more with redshifts above 6.5. Future surveys such as \textit{Euclid} will further increase the number of known quasars above redshift 7 \citep{euclid2019}. It is therefore relevant to  make predictions for the effective optical depth distributions beyond redshift 6 that may be measured from future observations of such quasars. As discussed in Section \ref{sec:sims}, we have modelled here two extremes in reionization history, and emphasise again that the High-$\tau_{\rm CMB}$ model is already disfavoured by the CMB. It is still interesting however to see what constraints can be obtained from future analyses of absorption line spectra. We demonstrate again the differences between the reionization histories of our three models in Figure \ref{fig:lc}, where we plot the evolution in \ion{H}{i} fraction in redshift along a lightcone through the simulation volume for each model. The ionization state of the gas is relatively similar below redshift 6 in all models, where they are constrained by the Ly$\alpha$ forest measurements. At higher redshifts, however, the contrast between both of the Low-$\tau_{\rm CMB}$ models and the High-$\tau_{\rm CMB}$ model is very striking as the differences between the midpoint of reionization in the two models become more apparent. The sizes of ionized bubbles in the High-$\tau_{\rm CMB}$ model are much larger than in both of the Low-$\tau_{\rm CMB}$ models, due to the higher ionizing emissivity assigned to each halo at early times (top left panel of Figure \ref{fig:history}).

Although there are clearly large reservoirs of ionized gas out to redshift 6.5 in all models, the \ion{H}{i} fraction inside the ionized bubbles increases with increasing redshift, as the  photoionization rate in the bubbles is lower before they percolate. This means that the incidence of opaque sightlines will increase in all models. Pushing  Ly$\alpha$ forest measurements to higher redshifts with current facilities will therefore be challenging, as the observations will become dominated by the noise of the sky background. We nevertheless show the evolution of the Ly$\alpha$ optical depths out to redshift 7.1 in the three models in the top panel of Figure \ref{fig:highz}. The evolution of the CDFs for the Low-$\tau_{\rm CMB}$ and Hot Low-$\tau_{\rm CMB}$ models are very similar, but the difference between those CDFs and the CDF of the High-$\tau_{\rm CMB}$ model grows with increasing redshift. Indeed, the  the High-$\tau_{\rm CMB}$ model predicts that there should be more than a factor 2 more sightlines with $\tau_{\rm eff, Ly\alpha} \gtrsim 8$ at $z=6.7$. The difference between the models however is close to the scatter that is expected from cosmic variance in measurements from a limited number of sightlines (the 15$^{\rm th}$/85$^{\rm th}$ percentiles of 1000 CDFs constructed using 10 sightlines is shown for the Low-$\tau_{\rm CMB}$ model in the blue shaded region, and for the High-$\tau_{\rm CMB}$ model in the yellow shaded region). The situation will be further complicated by the fact that any measurements of such high effective optical depths will likely be upper limits only.

\begin{figure*}
\includegraphics[width=2\columnwidth]{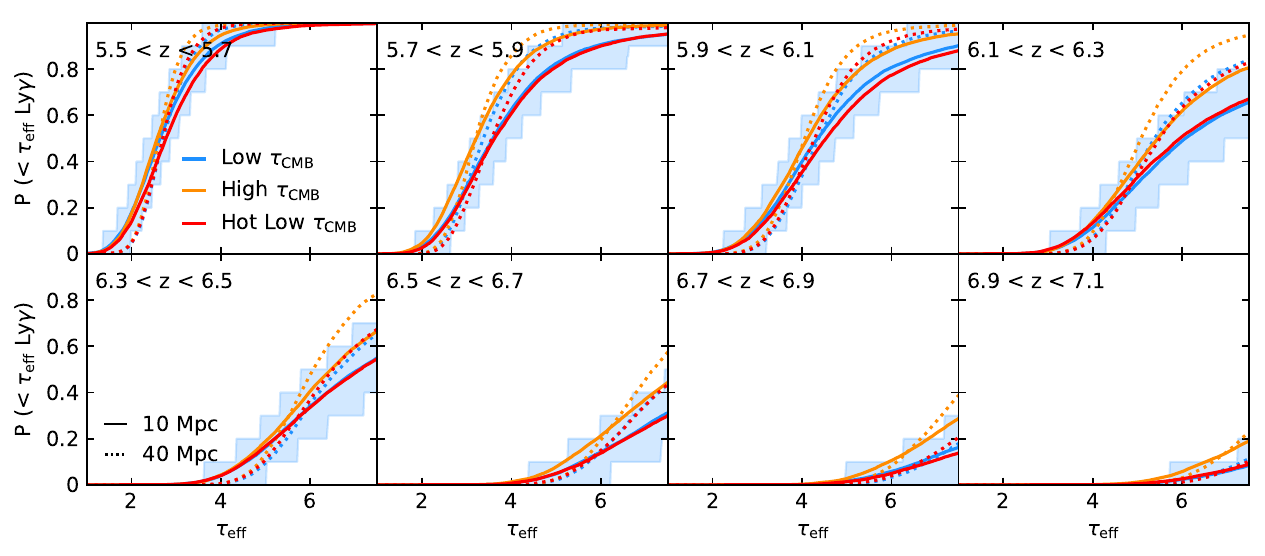}
    \caption{Cumulative distribution of the effective optical depth of the Ly$\gamma$ forest measured along skewers of length 10 Mpc (solid lines) and 40 Mpc (dotted lines). The pathlength accessible to observing the Ly$\gamma$ forest is not long enough to measure along 40 Mpc skewers, but we show it here for comparison with the Ly$\alpha$ and Ly$\beta$ CDFs presented elsewhere in this paper. The blue shaded region accounts for the effect of cosmic variance in the Low-$\tau_{\rm CMB}$ model: the shaded region spans the 15$^{\rm th}$ and 85$^{\rm th}$ percentile range for 1000 CDFs calculated assuming a future compilation of 10 measurements in each bin. The strength of this effect is similar for all models.}
    \label{fig:lyg}
\end{figure*}

A more promising avenue is to measure the evolution of the effective optical depths of the Ly$\beta$ forest, which we show in the bottom panel of Figure \ref{fig:highz}. Again, the difference between the Low-$\tau_{\rm CMB}$ and High-$\tau_{\rm CMB}$ models increases with increasing redshift. In this case, above $z=6.3$, the difference between the Low-$\tau_{\rm CMB}$ and High-$\tau_{\rm CMB}$ models is noticeably larger than the 15$^{\rm th}$/85$^{\rm th}$ percentile scatter predicted from cosmic variance (compare blue and yellow shaded regions). We also find that there is already a significant difference in the models below $\tau_{\rm eff, Ly\beta} \sim 5$, which is perhaps detectable in high signal-to-noise spectra. We therefore suggest that observations of the Ly$\beta$ forest out to high redshift can be used to constrain the evolution of the ionized fraction of the IGM out to $z\sim 7$, in the same way that the Ly$\alpha$ forest has been used to fix the ionization history below $z \sim 6$. If the midpoint of reionization is close to $z\sim 7$, as suggested by the CMB and estimates from quasar damping wings \citep{davies2018}, then measurements of the effective optical depth of the  Ly$\beta$ forest out past $z=6.5$ would allow us to constrain \textit{the entire second half of reionization} from the Ly$\alpha$ and Ly$\beta$ forests alone.

Further information could in principle be obtained from the evolution of the Lyman-$\gamma$ (Ly$\gamma$) effective optical depth, although now the foregrounds of both the lower redshift Ly$\alpha$ and Ly$\beta$ absorption must be accounted for. The pathlength that is accessible by higher order Lyman series lines also becomes increasingly short for higher order lines, so measurements along sightlines towards quasars that have small proximity zones will be more favourable \citep[e.g.,][]{fan2006}. We estimate that, for a given sightline, 10 comoving Mpc of the Ly$\gamma$ forest could be observable (where we have assumed that 5000 km s$^{-1}$ of the Ly$\gamma$ forest at the high redshift end should be removed due to contribution from the proximity zone, and 1000 km s$^{-1}$ should be removed at the low redshift end to avoid the Lyman-$\delta$ emission peak). In Figure \ref{fig:lyg}, we present our prediction for the CDFs that would be observed. Unlike the Ly$\beta$ CDFs, we do not see a strong difference between the models with different midpoints of reionization. It seems that this is due to the contamination by the lower redshift Ly$\alpha$ and Ly$\beta$ forests rather than the shorter pathlength we measure Ly$\gamma$ along here. We demonstrate this by showing that we recover very similar results for the CDFs if we imagine we could measure the mean flux along 40 Mpc segments (as we have done elsewhere in this paper). As expected, the 40 Mpc CDFs are somewhat less broad (as the variations due to the density field or residual neutral islands are not felt so strongly), but again there is not a strong difference between the three models we present. We therefore conclude that the best chance of probing deeper into reionization with quasar absorption lines is through observations of the Ly$\beta$ forest.

\section{Conclusions}
\label{sec:conclusions}

We have presented here a study of the evolution of the Ly$\beta$ forest in cosmological radiative transfer simulations of models with reionization ending at $z=5.2-5.3$. The models we test have different midpoints of reionization and gas temperatures. We find that all of our models agree reasonably well with the observations of \citet{eilers2019} over the redshift range $5.5<z<6.1$, which corresponds to the tail-end of reionization in our models. The main discrepancy between the simulations and observations is in the incidence rate of the sightlines with the highest Ly$\beta$ optical depths in the redshift bin $5.7 < z <5.9$, however our simulations do reproduce several sightlines that are in agreement with observations. We further note that the sightline that these observations are measured along is the long absorption trough of ULAS J0148+0600, which we have already shown to be rare in radiative transfer simulations such as these. 

We conclude that the late-end reionization scenario proposed in \citet{kulkarni2019} to explain the properties of the high redshift Ly$\alpha$ forest is also in good agreement with observations of the Ly$\beta$ forest over the same redshift range. We therefore find there is no ``anomaly'' in the opacity of the Ly$\alpha$ and Ly$\beta$ forests at high redshift as has previously been suggested, and that there is no need to invoke an inverted temperature-density relation in the IGM above $z>5.5$ to explain the observations. Good agreement between models with reionization ending late and the observational data seems however to require proper modelling of radiative transfer effects, as the semi-numerical method of \citet{nasir2019} does not seem to produce as many sightlines that are opaque in Ly$\beta$ in the redshift range $5.7<z<5.9$ as in the radiative transfer simulations we present here.

We find that the current published effective optical depths do not yet discriminate between our models with different midpoints of reionization. We show that future measurements of the effective optical depth of the Ly$\beta$ forest pushing out to higher redshifts will differentiate between reionization histories with redshifts of the midpoint of reionization differing by $\Delta z > 1.5$. There are already enough $z > 6.5$ quasars known to measure the evolution of the Ly$\beta$ forest out towards $z=7$, so all that remains is to take spectra of high enough quality to provide reasonable constraints. An ongoing ESO large programme taking high signal-to-noise spectra of a sample of $z>5.8$ quasars with VLT/X-SHOOTER (PI: V. D'Odorico) is a very promising step in this direction.

\section*{Acknowledgements}

This work was performed using the Cambridge Service for Data Driven Discovery (CSD3), part of which is operated by the University of Cambridge Research Computing on behalf of the STFC DiRAC HPC Facility (www.dirac.ac.uk). The DiRAC component of CSD3 was funded by BEIS capital funding via STFC capital grants ST/P002307/1 and ST/R002452/1 and STFC operations grant ST/R00689X/1. This work further used the DiRAC@Durham facility managed by the Institute for Computational Cosmology on behalf of the STFC DiRAC HPC Facility. The equipment was funded by BEIS capital funding via STFC capital grants ST/P002293/1 and ST/R002371/1, Durham University and STFC operations grant ST/R000832/1. DiRAC is part of the National e-Infrastructure. This work was supported by the ERC Advanced Grant 320596 ``The Emergence of Structure During the Epoch of Reionization''.

\section*{Data Availability}

The data underlying this article are available at 10.5281/zenodo.3921822.

\bibliographystyle{mnras}
\bibliography{ref} 

\appendix

\section{Forward modelling of spectral noise}

\begin{figure*}
\includegraphics[width=2\columnwidth]{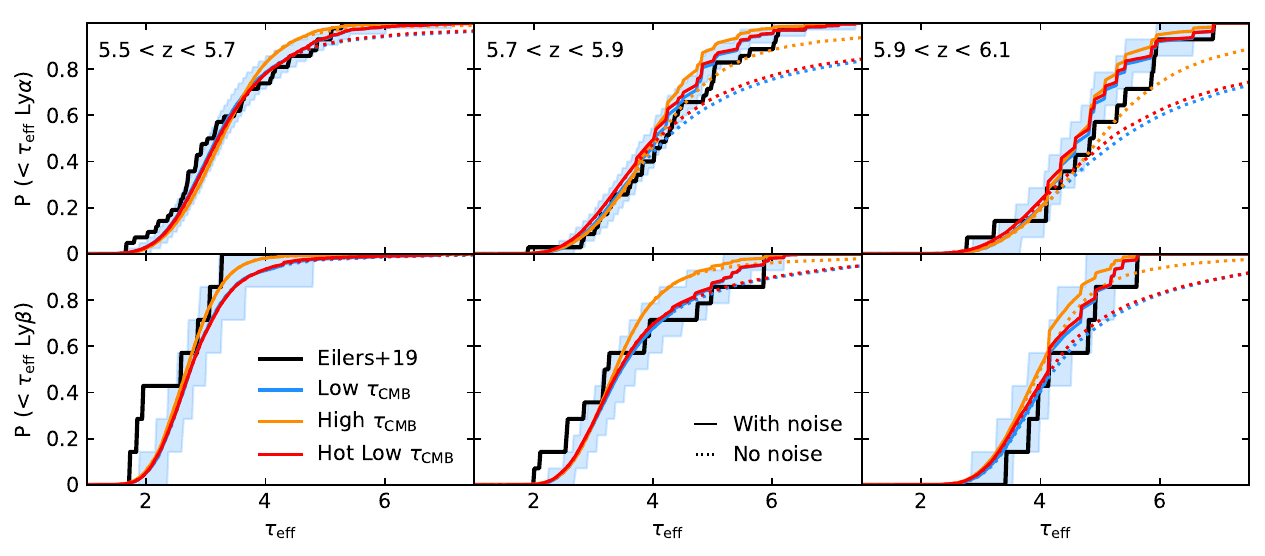}
    \caption{Cumulative distribution of the effective optical depth of the Ly$\alpha$ forest (top row) and Ly$\beta$ forest (bottom row) measured along skewers of length 40 Mpc. We show three different redshift intervals: $5.5 < z < 5.7$ (left), $5.7 < z < 5.9$ (middle) and $5.9 < z < 6.1$ (right). The coloured lines show the median CDFs for each model (blue for Low-$\tau_{\rm CMB}$ model, orange for High-$\tau_{\rm CMB}$ and red for Hot Low-$\tau_{\rm CMB}$). The blue shaded region accounts for the effect of cosmic variance in the Low-$\tau_{\rm CMB}$ model: the shaded region spans the 15$^{\rm th}$ and 85$^{\rm th}$ percentile range for 1000 CDFs calculated using the same number of sightlines as the \citet{eilers2019} compilation. The strength of this effect is similar for all models. The solid lines are CDFs constructed using sightlines with forward modelled spectral noise. The dashed lines show the case with no noise. The black lines show the cumulative distributions calculated from the \citet{eilers2019} sample, counting the lower limits on $\tau_{\rm eff}$ as detections.}
    \label{fig:cdfsnoise}
\end{figure*}

Throughout this paper, we have determined whether the opacity distribution of our simulated spectra are a compatible with observations by seeing if they lie within the ``optimistic'' and ``pessimistic'' predictions of the data as in \citet{bosman2018}. In this scenario, if the mean flux of a sightline is below the noise, then its mean flux is assumed to be at most twice the noise. The other extreme is to assume that there is no transmitted flux at all. Rather than comparing the simulations against these limits, one can also forward model the spectral noise as in \citet{eilers2019}. In this appendix, we compare our simulations against observations using the forward modelling technique.

We construct 1000 mock surveys, drawing random sightlines from our simulations and adding noise appropriate to the uncertainties of the \citet{eilers2019} observations. As with the observed data, we note the simulated sightlines which have a mean flux $\langle F \rangle < 2 \sigma_{\rm obs}$, where $\sigma_{\rm obs}$ is the uncertainty in the mean flux along a given sightline. In these cases we assume $\langle F \rangle = 2 \sigma_{\rm obs}$ for this sightline. Otherwise we take the mean flux calculated for that sightline.

Our results are shown in Figure \ref{fig:cdfsnoise}. As in \citet{eilers2019}, we find that forward modelling the noise in the simulated data narrows the opacity CDFs. However, we find consistent results with the analysis presented in section \ref{sec:cdfs}. This is in contrast to \citet{eilers2019}, who found that accounting for this spectral noise put their models in tension with the observations. We attribute the difference here to the fact that the opacity CDFs produced in these late reionization models are significantly broader than the UV and temperature fluctuation models before any noise is added.

\bsp	
\label{lastpage}
\end{document}